%% file: main.tex
\documentclass[letterpaper,twocolumn,10pt]{article}
\usepackage{usenix}

\title{DNS-based Ingress Load Balancing:\\An Experimental Evaluation}

\newcommand{\supsym}[1]{\raisebox{4pt}{{\footnotesize #1}}}
\newcommand{\gt}{\supsym{$\dag$}}
\newcommand{\jt}{\supsym{$\ddag$}}

\date{}
\author{
{\rm Partha Kanuparthy~\gt, Warren Matthews~\jt, Constantine Dovrolis~\gt}\\
\begin{small}
{\gt~Georgia Institute of Technology~~~~ \jt~JANET}
\end{small}
}

\usepackage{color}
\usepackage{subfigure}
\usepackage{float}
\usepackage{graphicx}
\usepackage{alltt}
\usepackage{multirow}
\usepackage{psfrag}
\usepackage{balance}

\usepackage{amsmath}
\usepackage{amsfonts}
\DeclareSymbolFont{AMSa}{U}{msa}{m}{n}
\DeclareMathSymbol{\rightsquigarrow}{\mathrel}{AMSa}{"20}
\usepackage{url}

\renewenvironment{thebibliography}[1]{%
  \begin{oldthebibliography}{#1}%
  \setlength{\parskip}{0ex}%
  \setlength{\itemsep}{0ex}%
}{\end{oldthebibliography}}

\begin{document}

\maketitle

\input{abstract.tex}

%\begin{IEEEkeywords}
%Load balancing, traffic engineering, DNS, measurement.
%\end{IEEEkeywords}

\input{intro.tex}

\input{gtlib.tex}

\input{oursystem.tex}

\input{lberror.tex}

\input{mbscheme.tex}

\balance
\input{conclusion.tex}

%\begin{scriptsize}
%\begin{small}
\bibliographystyle{plain}
\bibliography{ref}
%\bibliographystyle{ref}
%\end{small}
%\end{scriptsize}

\end{document}

%% file: abstract.tex
\begin{abstract}
Multihomed services can load-balance their incoming connection
requests using DNS, resolving the name of the server with different
addresses depending on the link load that corresponds to each address.
Previous work has studied a number of problems with this approach, e.g., 
due to Time-to-Live duration violations and client proximity to local DNS servers. In
this paper, we experimentally evaluate a DNS-based ingress traffic engineering
system that we deployed at Georgia Tech. Our objective is to
understand whether simple and robust load balancing algorithms can be
accurate in practice, despite aforementioned problems with DNS-based load
balancing methods. In particular, we examine the impact of various system
parameters and of the main workload characteristics. We show that
a window-based measurement scheme can be fairly accurate in practice,
as long as its window duration has been appropriately configured.
\end{abstract}

%% file: intro.tex
\section{Introduction}
\label{sec:intro}

As cloud services and content delivery become increasingly
ubiquitous, multihoming is turning to be an integral part of
network infrastructure, to distribute load and for failover.
A recent study \cite{dhamdheretwelve}, for example, found that the average AS degree increased
by one link per AS over the last 12 years.

\emph{Ingress Traffic Engineering (ITE)} aims to select
an incoming link among a set of possible links
for the communication between a multihomed data center
(or server farm) and its client population.
The primary objective of ITE is to load-balance 
the incoming traffic to (and consequently, the outgoing 
traffic from) the data center. A secondary objective is to choose a
better path (e.g., minimum delay or maximum available bandwidth) 
for each client.  Typically, however, content providers are primarily
interested in avoiding congestion in their own access links,
and so the load-balancing objective is their primary concern. 

Networks today employ two common
approaches to do ITE. The first approach relies on BGP
and selective prefix advertisements \cite{10.1109/ICDCS.2008.112,quoitin2003ite}.
This approach can balance load at the level of 
IP address blocks, not client networks, it creates
BGP churn at the Internet core, and it may be subject
to BGP route dampening and convergence delays. 
The second approach uses the DNS infrastructure
to dynamically select one of $k$ IP addresses to resolve the
server's name, where $k$ is the number of incoming links. 
The DNS-based solution is becoming
increasingly popular, since it
can balance the server's load at the granularity of
individual DNS requests, and it does not require 
the content provider to use BGP.

\begin{figure*}
\begin{center}
 \includegraphics[width=0.7\textwidth]{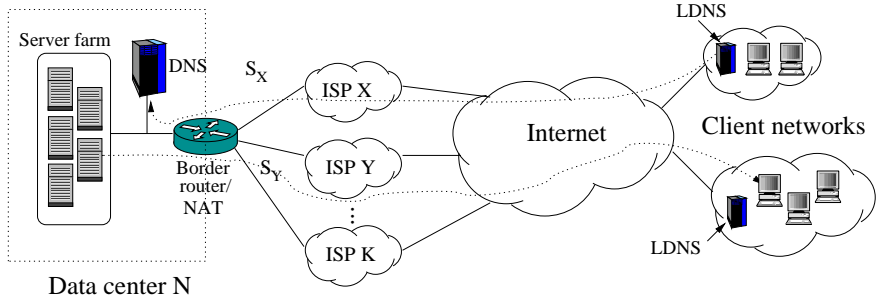}
\caption{\label{fig:dns-ite}The DNS-NAT ingress traffic
engineering architecture.}
\end{center}
\end{figure*}

The most commonly deployed scenario of DNS-based 
ITE is the DNS-NAT architecture shown in Figure~\ref{fig:dns-ite}.
Consider a multihomed network $N$ that uses IP
addresses from two ISPs $X$ and $Y$. 
It is easy to control the assignment of
outgoing connections (initiated from the data center servers) to
egress links. To control the egress link of traffic in 
client-initiated connections, however, is more challenging. 
One way to do so it is to rely on NATs and DNS, as follows.
A server $S$ in $N$ is statically NATed with two IP addresses 
$S_X$ and $S_Y$, from $X$ and $Y$ respectively. Clients requesting
content from server $S$ first resolve its hostname, and 
then establish a TCP connection to $S$. Network $N$ runs
an authoritative DNS server for the domain name of $S$,
which resolves each incoming DNS request from a client 
{\em Local-DNS} (LDNS) server with either $S_X$ or $S_Y$.
Thus, traffic between clients and the data center is 
routed on a per-LDNS basis through ISPs $X$ or $Y$.
Of course the same approach can be followed in the 
case of more than two upstream ISPs.

The ITE method presents some hard challenges.
First, the traffic that follows a DNS name resolution
can consist of multiple TCP connections, due to DNS caching
at the client. 
Second, multiple client sessions can follow the same LDNS request, 
due to caching at LDNS servers.
Consequently, at DNS resolution time, we do not know the magnitude
or duration of the traffic that follows each DNS request. 
Further, that traffic cannot be partitioned between different
upstream ISPs. 
Third, TCP can cause significant variations in the throughput
of the incoming load, making it harder to predict the incoming
traffic on each interface.
Fourth, there can be a significant delay between a DNS request 
(and the associated load balancing decision) and the arrival of the actual 
traffic that corresponds to that DNS request.

Previous work has studied a number of issues with DNS-based
ITE approaches.
First, short advertised DNS Time-To-Live (TTL) durations are not always 
honored by remote DNS servers or clients \cite{pang2004rdb}.
Second, it is not possible to migrate an ongoing TCP connection from one link 
(i.e., server address) to another.
Third, clients are not always near their 
LDNS servers, which can affect accuracy of ITE methods
that attempt to select the best path for each client \cite{mao2002pae}.

Theoretical studies have shown the effectiveness of
randomized load balancing \cite{mitzenmacher2001ptc}, and issues with
stale measurements \cite{mitzenmacher2000uoi}. It is known that
prior information about incoming jobs can help \cite{dahlin2000isl,harcholbalter1999cta}.
In ITE, however, it is not always feasible to predict the 
arrivals or size of client DNS requests. In addition, empirical
studies of ITE that compliment our work include 
\cite{cardellini1999dlb,guo2004,Akella:usenix2004}.

%In this paper, we first examine some relevant traffic characteristics 
%of a popular content distribution service at Georgia Tech.
%This site (\url{http://www.gtlib.gatech.edu})
%provides regular updates and releases of several Linux and 
%BSD distributions as well as application software to thousands of 
%Internet users.
In this paper, we evaluate a DNS-based ingress traffic
engineering system that we deployed at Georgia Tech.
Our objective is to understand whether a simple but robust load balancing 
algorithm can be accurate in practice, despite all known problems with 
DNS-based load balancing methods. 
%We find that load balancing performance depends on several factors, 
%including client request rates, number of clients using an LDNS, transfer
%sizes, TTL violations by remote networks, and traffic 
%variability. We show that load balancing errors can
%be significantly reduced when using a window-based measurement algorithm, 
%over a round-robin scheme.
We also examine the impact of the measurement window duration on 
load balancing accuracy.

The rest of the paper is organized as follows. 
%We start with some observations about the workload of a popular content 
%provider  in Section~\ref{sec:gtlib}.
Section~\ref{sec:oursystem} describes our ITE prototype and 
workload.  
Section~\ref{sec:perf} shows experimental results 
for the impact of key system and workload characteristics on
load balancing accuracy. In Section~\ref{sec:mbscheme}, we
analyze the impact of window duration on load balancing using a 
history-based measurement algorithm.
%We conclude by discussing related work in Section~\ref{sec:relwork}.

%% file: gtlib.tex
\section{Workload Characteristics of a Content Provider}
\label{sec:gtlib}

In this section, we present measurements from the GTLIB 
content distribution service at Georgia Tech. This service
is provided by four servers whose hostname, \url{www.gtlib.gatech.edu},
is resolved in a round-robin fashion by two authoritative DNS
servers.  We have collected full
Netflow traces from the Georgia Tech campus border router.
In parallel, we run tcpdump at the authoritative DNS servers of
the previous site. The measurements were performed for a  
duration of 24 hours starting on 10th April 2008 at 9pm.
We analyzed the collected traces (the capture clocks
were kept synchronized via NTP) to first understand whether 
remote LDNS servers honor the advertised DNS TTL of 8 hours,
and second, to characterize the traffic workload in terms
of bytes per client and clients per LDNS.

\begin{figure*}
%\begin{centering}
\subfigure[Request inter-arrivals from same LDNS]{\includegraphics[width=0.33\textwidth]{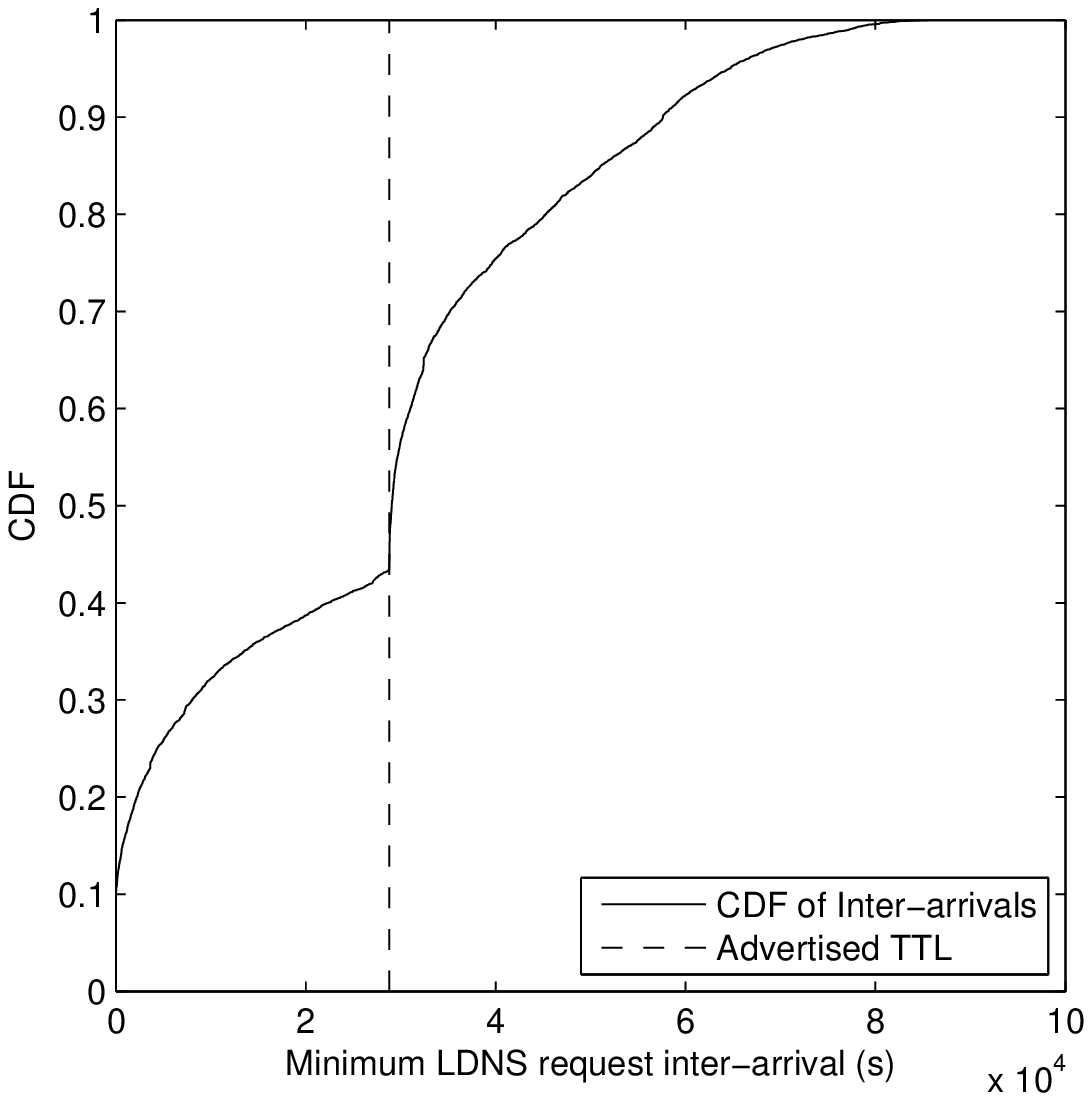}\label{fig:gtlib-ttl}}
\subfigure[Bytes per client]{\includegraphics[width=0.35\textwidth]{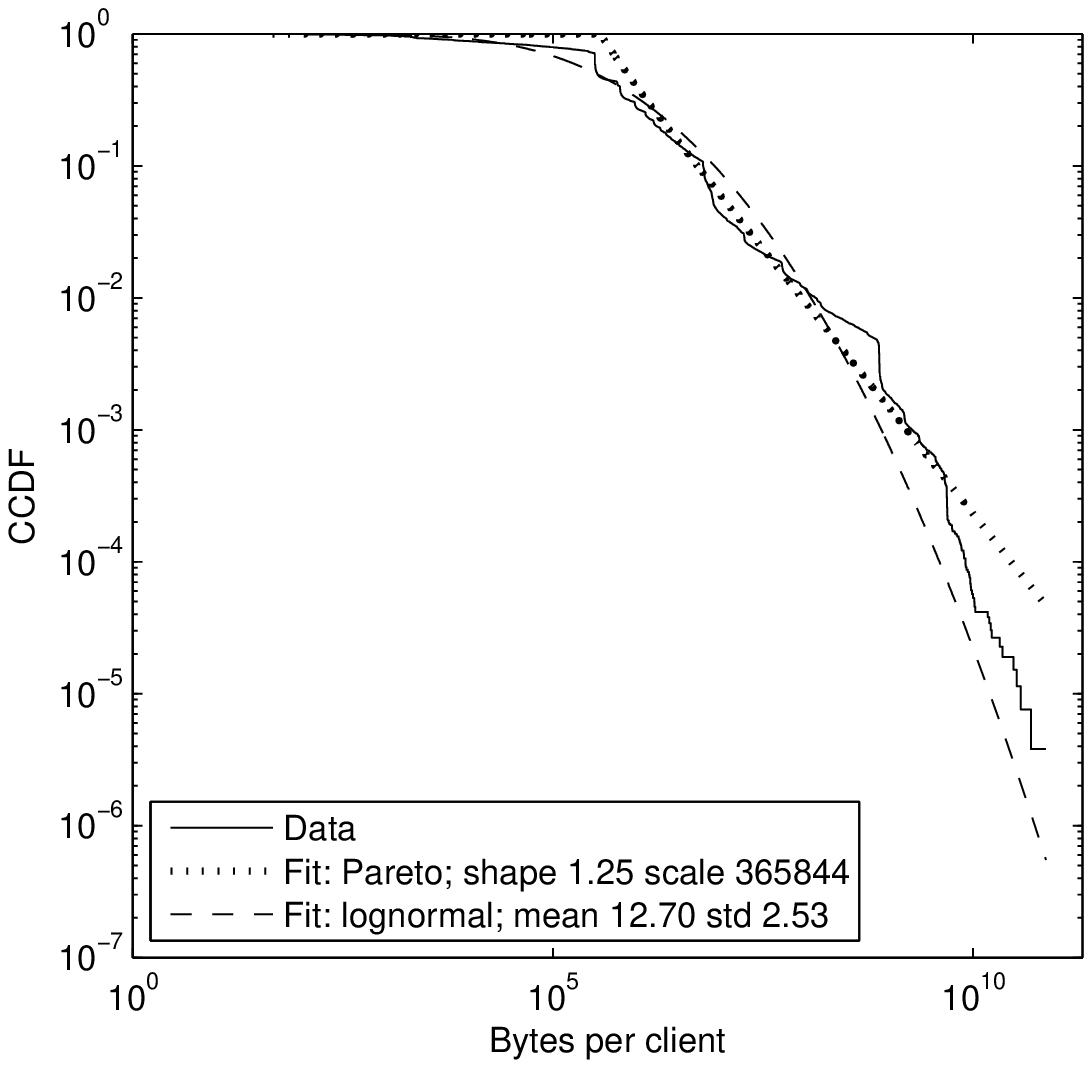}\label{fig:gtlib-filesize}}
\subfigure[Clients per LDNS]{\includegraphics[width=0.35\textwidth]{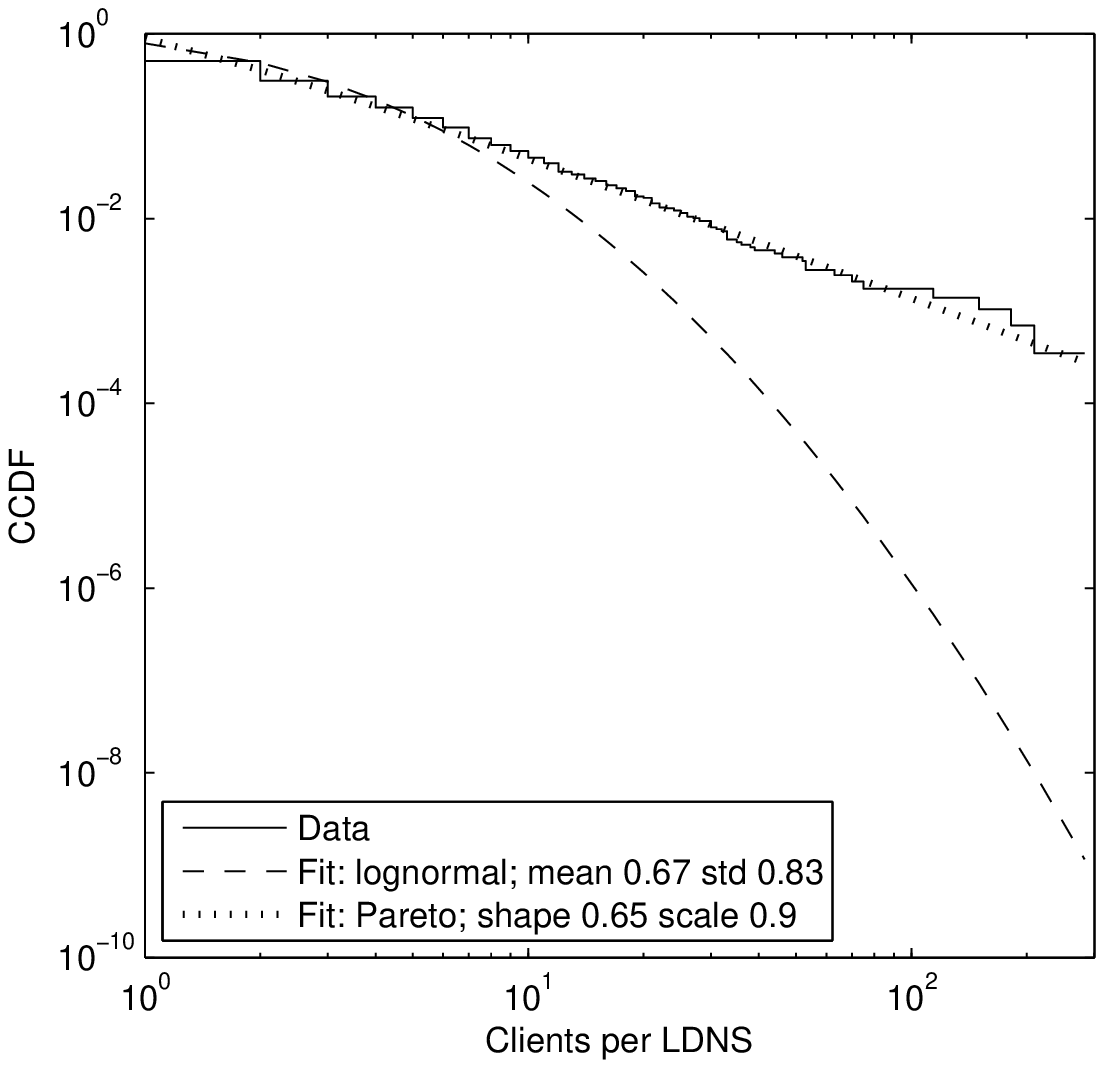}\label{fig:gtlib-hiddenclients}} 
%\end{centering}
\caption{Workload characteristics from GTLIB content distribution service.\label{fig:gtlib-cdfs}}
\end{figure*}

\textbf{How many LDNS servers honor the advertised TTL?}
We first look at the distribution of the minimum 
inter-arrivals from remote LDNS servers.
The A records for \url{www.gtlib.gatech.edu} pointed to
\url{128.61.111.[9-12]}, and the advertised TTL was 8 
hours. There were about 46,400
resolutions of type A. Figure~\ref{fig:gtlib-ttl} shows
the distribution of minimum inter-arrival times for each
LDNS server that contacted our authoritative servers. The figure
shows that for around 60\% of the remote LDNS servers, we received
requests at most once per 8 hours. We can say
that these LDNS servers either follow the advertised TTL and/or
have a lower client resolution request rate than one per 8 hours. 
For the remaining 40\% of the LDNS servers, we expect that they 
either use a TTL of less than 8 hours, or that they do
not do caching. In other words, these LDNS servers violated the
advertised TTL of 8 hours. Moreover, about 10\% of them have a very
short minimum inter-arrival period, which implies that they may not
be doing any caching.

\textbf{How many clients correspond to each LDNS server?}
The challenge in this measurement is how to associate an LDNS request
with all subsequent arriving connections from clients that have
used that LDNS request to resolve the server's name.  
We use a simple approach to do this correlation, based on the
hypothesis that an LDNS server and its associated clients 
belong to the same Autonomous System, and thus their addresses would both
have the same BGP AS-Origin attribute (even though they
often do {\em not} belong to the same IP address prefix).
First, for each client request from an address A, 
we find earlier DNS requests from LDNS servers that 
belong to the same Autonomous System that advertises address A.
If there are no such LDNS servers, or if there are 
multiple such servers, we ignore that client request.
Otherwise, we associate that client request with 
the most recent DNS request from the corresponding LDNS server.
Note that we 
ignore clients which resolved the GTLIB hostname before the start
of our 24-hour dataset. We also ignore clients that round-robin
between DNS servers in their resolver configurations. Most OSes 
either do not support this feature or disable it by default.
Using the previous approach, we found
that we can associate 92\% of the client sessions with
an LDNS entry from the same origin-AS with the client. 
In total, we identified 2864 unique LDNS servers
that we could associate with specific client requests. 
Figure~\ref{fig:gtlib-hiddenclients} shows the distribution of 
the number of clients using a given LDNS.
We also show regression curves for Pareto and lognormal distributions. 
Note that the Pareto distribution is a better fit, showing 
that the number of clients that correspond to each LDNS is highly skewed.

\textbf{How many bytes were received by each client?}
Figure~{\ref{fig:gtlib-filesize} shows the distribution of
bytes that the GTLIB servers sent/received from each unique client address 
in the course of the 24-hour trace.
Note that the lognormal distribution is a better fit to the measured data.

We use the previous GTLIB observations to emulate realistic 
workload in the experimental evaluation of a DNS-based ITE system,
described in the next section.

%% file: oursystem.tex
\section{System Implementation and Deployment}
\label{sec:oursystem}

In this section, we describe our DNS-ITE prototype.
The Georgia Tech campus network has several commercial
and research providers. 
We multihome our server using two IP addresses
which are advertised to the Internet through Qwest, Cogent, and Internet2.
The server's hostname is dynamically resolved by our DNS server 
to one of the two addresses.
%The server's hostname,
%\url{beast.ingress-expt.rnoc.gatech.edu},
%is dynamically resolved by our DNS server to either 
%\url{143.215.250.110} or \url{128.61.113.255}.

\textbf{Load-balancing Algorithms:}
We perform load balancing using two algorithms: 
(1) round-robin (RR) and (2) measurement-based (MB). 
The RR scheme selects the server's IP address 
in a round-robin fashion.
The MB algorithm uses recent history of ingress 
and egress traffic at the interfaces of our server
to make a load-based decision about the next name resolution. 
Our goal is to understand how simple round-robin and history-based
schemes work in practice.

\textbf{Implementation:}
Our ITE system consists of two processes running on the same host, 
the DNS process and a monitoring process.
The DNS process is a non-blocking and concurrent,
non-recursive authoritative nameserver,
which serves LDNS resolution requests for our domain.
The DNS process communicates with the monitoring process
to get traffic measurements.
This communication is done using
shared memory; we have also tested an RPC-based
mechanism to run the two processes on different hosts.
The monitoring process measures aggregate traffic 
utilization at the two interfaces of the server. We
measure the load on each interface using a \emph{sliding window} of 
length $W$=$n w$ seconds, which consists of $n$ small windows of length
$w$. The sliding window moves in steps of $w$ seconds. In our 
implementation, we set $w=100$ms. 
Our prototype can be extended to other 
traffic measurement methods such as Netflow,
and to multiple content servers.

\textbf{Server characteristics:}
We run an Apache server on Linux 
serving content of client requested size over HTTP.
The server, DNS and monitoring processes run on a 
2GHz hyper threading-enabled Xeon with 1GB
physical memory. At peak experiment loads of 5-10 Mbps
traffic, the CPU usage does not exceed 20\%. 

\textbf{Workload:}
We emulate a realistic workload by using
40 clients on PlanetLab and 6 clients on RON networks.
Some of our workload parameters
are drawn from observations of Georgia Tech's GTLIB content
distribution mirror.
We choose these nodes such that they
use different LDNS servers (i.e. have disjoint 
resolver configurations). Further, we pick
LDNS servers that follow the advertised TTL (so that, for example,
we can emulate LDNS servers using a minimum
TTL).  The duration of each experiment
is 10 minutes.
%\footnote{We ensure that we do not exceed the bandwidth 
%limits of PlanetLab hosts during our experiments.}

The traffic model that we emulate is described next.
First, 40\% of LDNS servers do not follow the
nominal TTL of 15 seconds - we advertise a 15s TTL to those servers.
For these LDNS servers, we advertise a TTL that is 
uniformly distributed in $[5,600]$s. We always advertise
the same TTL to a given LDNS server. The fraction of TTL
violations is based on DNS logs from GTLIB.
Second, clients follow a closed-loop (or {\em interactive}) arrival model, 
in which they download a file over TCP, 
sleep for some time, and then repeat this process.
Unless stated otherwise, 
the sleep times are exponentially distributed with a mean of 35 seconds. 
Third, clients download lognormally distributed file sizes
with a mean of 225KB, based on a 24-hour Netflow data from GTLIB. 
However, we had to truncate 
the size distribution to 625KB to avoid exceeding PlanetLab
byte limits that could trigger rate-limiting.
Fourth, there can be many clients behind an LDNS.
We refer to these clients as \emph{hidden clients}.
We emulate them by spawning multiple simultaneous processes
on the same client host. The number of hidden clients on each host is drawn from
the uniform distribution $[1,5]$.
%Our goal here is to study
%the effect of hidden clients on load balancing.
Finally, the emulated clients have diverse path characteristics 
(RTT and available bandwidth) to our server, as it would
also happen in a real content provider.

%% file: lberror.tex
\section{DNS-ITE Performance}
\label{sec:perf}

The accuracy of any load balancing scheme 
depends on the job size granularity 
at which we can ``route'' jobs to servers.
In the context of DNS-ITE, this granularity is the  
the number of bytes that correspond to each LDNS 
request. If each LDNS request was
followed by only few bytes worth of load, we would be
able to achieve much more accurate load balancing than
if each LDNS request was followed by a large
and long file transfer. 
In this section, we start with a model that describes
the factors on which the DNS-ITE load balancing granularity
depends on. We then empirically evaluate how
the accuracy of a round-robin load balancer depends on these parameters.

Consider $n$ clients behind each LDNS server. Suppose
each client downloads $s$ bytes from the
server in each connection. Let the arrival
rate of connections per client be $r$, and
the arrival rate of DNS requests from each LDNS 
be $\lambda$. 
If a remote LDNS server uses caching, the TTL that it follows is $T$ seconds.
Then, 
\[\lambda = \begin{cases}
n r & \text{if LDNS is non-caching}\\
\min \{ n r, \frac{1}{T} \} & \text{LDNS caches with TTL $T$}\end{cases}\]
The traffic rate that corresponds to an LDNS server is given by
$R = n r s$ (bps). The granularity in which we can balance the
arriving traffic is:
\begin{equation}
\frac{R}{\lambda} = \begin{cases}
s & \text{if $i$ is non-caching, or } n r < \frac{1}{T}\\
n r s T & \text{otherwise}\end{cases}
\label{eq:traffic-per-ldnsreq}
\end{equation}
Next, we quantify the effect of the parameters that control this 
ratio $R/\lambda$
on the load balancing accuracy that we can achieve with
a simple RR algorithm.

\textbf{The error metric:}
We quantify the load balancing accuracy in terms of the
\emph{relative difference} between the utilization of the two links.
measured in an {\em averaging timescale} of $I$ seconds. 
More precisely, we measure the traffic utilization (in bps)
$U_{I,1}(t)$ and $U_{I,2}(t)$ of the two links at our server in
a sliding window of length $I$ that starts at time $t$. 
Under perfect load balancing
conditions, the load on each link during the interval $(t,t+I)$ 
should be $[U_{I,1}(t) + U_{I,2}(t)]/2$.
The load balancing error $\epsilon$  
is defined as: 
\[
\epsilon_I(t) = \frac{| U_{I,1}(t) - U_{I,2}(t) |}{U_{I,1}(t) + U_{I,2}(t)}
\]
We re-compute $\epsilon_I(t)$ every one second.

\begin{figure}
%\vspace{-30pt}
\begin{center}
\includegraphics[width=0.45\textwidth]{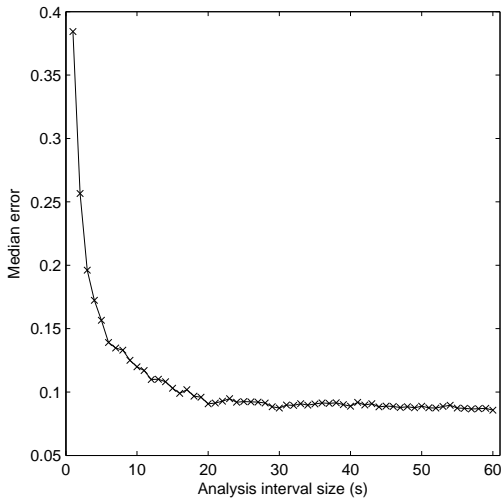}
\end{center}
%\vspace{-20pt}
\caption{Variation of median error with averaging timescale $I$.\label{fig:error-window}}
%\vspace{-10pt}
\end{figure}

Figure~\ref{fig:error-window} shows the impact of the
sliding window length $I$ on the median error $\epsilon_I(t)$,
for one of our experiments.
As expected, the load balancing error is higher as we
decrease $I$, as there are fewer arriving connection
requests in shorter intervals. 
Also note that the error metric tends to stabilize when 
$I$ is larger than about 15 seconds.
This shows that the RR load balancing algorithm is not
able to eliminate the relative error, even when we use
a significantly long (15s) averaging timescale. 
In the rest of this paper, we use $I$=20s.

\begin{figure}[t]
\begin{center}
\includegraphics[width=0.45\textwidth]{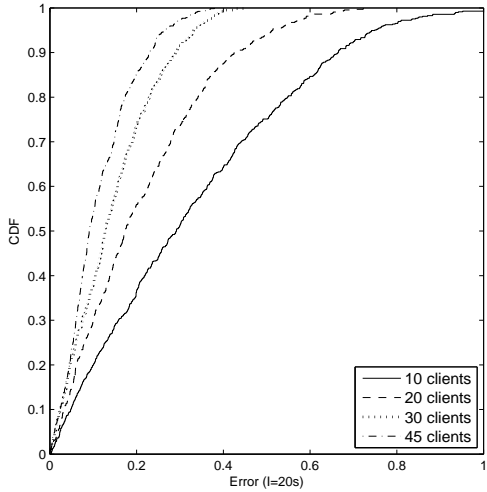}
\end{center}
\caption{Effect of aggregate load (RR).\label{fig:clientrate}}
\end{figure}

\textbf{Aggregate load:}
Our goal in this experiment is to examine the impact of
the aggregate traffic load on the relative error. 
We do so by varying the number of deployed LDNS servers
(and hence the number of active clients) from 10 to 45. 
Figure~\ref{fig:clientrate} shows the distribution of 
the relative error.
Note that the error decreases as the aggregate utilization
increases. 
As we decrease the number of LDNS servers and the associated clients,
the aggregate load drops, decreasing the frequency between 
arriving DNS requests. Thus, the load balancer has fewer
opportunities to distribute the arriving load between the
two links. 

\begin{figure}[t]
\begin{center}
\includegraphics[width=0.45\textwidth]{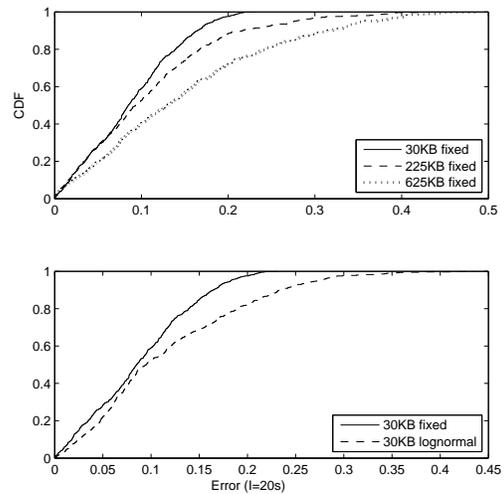}
\end{center}
\caption{Effect of file size (RR).\label{fig:filesize}}
\end{figure}

%\begin{figure*}
%\begin{minipage}[b]{0.48\linewidth}
%\begin{center}
%\includegraphics[width=1\textwidth]{figs/clientrate.eps}
%\end{center}
%%\vspace{-20pt}
%\caption{Effect of aggregate load (RR).\label{fig:clientrate}}
%\end{minipage}
%\begin{minipage}[b]{0.48\linewidth}
%\begin{center}
%\includegraphics[width=1\textwidth]{figs/filesize.eps}
%\end{center}
%%\vspace{-20pt}
%\caption{Effect of file size (RR).\label{fig:filesize}}
%\end{minipage}
%%\vspace{-10pt}
%\end{figure*}

\textbf{File size distribution:}
Here, we first examine the effect of the requested file size
and second, the effect of the significant variability in
the lognormal file size distribution compared to the case of 
constant file sizes. In order to keep the 
aggregate load fixed, we adjust the mean 
idle period between requests from each client. 
The top part of Figure~\ref{fig:filesize} shows 
effect of increasing the file size on the error 
distribution, when clients request fixed-sized files. 
Note that as we increase the file
size from 30KB to 625KB, the errors increase.
This is expected, because with smaller flows, 
our DNS server can take more frequent load balancing
decisions, amortizing the load between the two 
servers at a finer granularity. 

The bottom part of Figure~\ref{fig:filesize} shows the
differences between fixed-size transfers and 
lognormally-sized transfers.
We use a truncation size of 1MB for the latter.  
We see that the errors increase when the content size is 
heavy-tailed.
The main reason is that the round-robin scheduler
does not consider the current load on each link.
Thus, in a heterogeneous workload with transfers
of different sizes, it can happen that one link
receives several long transfers while another receives
mostly short transfers, causing periods of significant
imbalance. 
%\begin{figure}[!h]
%\begin{center}
%\includegraphics[width=0.4\textwidth]{figs/filesize.eps}
%\end{center}
%\caption{Effect of file size distribution (RR). \label{fig:filesize}}
%\end{figure}

\textbf{Advertised TTL:}
Next, we illustrate the effect of advertised TTL on
the load balancing accuracy. Equation~\ref{eq:traffic-per-ldnsreq}
shows that accuracy depends on the TTL $T$ that is advertised to LDNS 
servers, as long as the request rate from each LDNS $r n$ is
larger than $1/T$. 
When the request rate is less than that, we expect that
the load balancing error will not depend on TTL. 

\begin{figure}[t]
\begin{center}
\includegraphics[width=0.45\textwidth]{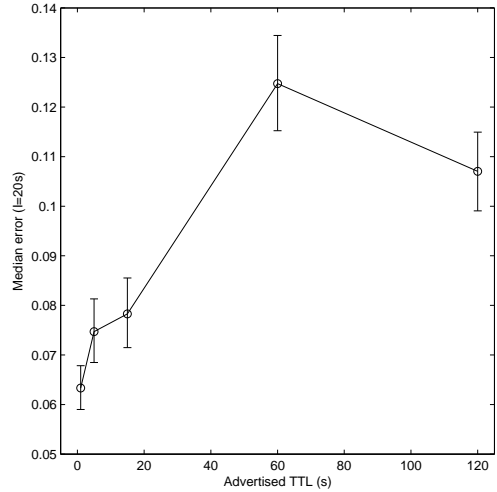}
\end{center}
\caption{Effect of advertised TTL (RR).\label{fig:ttl-effect}}
\end{figure}

We examined the impact of the advertised TTL on the median
of the relative error $\epsilon$ as follows. The 
client request rate $r$ is set to once per 35s. Hence, the
request inter-arrival $r n$ from an LDNS varies 
between 7s and 35s, depending on the number of clients per LDNS. 
Figure~\ref{fig:ttl-effect} shows the median error 
and Wilcoxon-based 99\% confidence intervals
for different advertised TTLs. Consistent with 
Equation~\ref{eq:traffic-per-ldnsreq}, when the
time period between successive DNS requests from the same LDNS
is less than the advertised TTL,
the error increases with the TTL (TTL values 1, 5, and 15 seconds).
For the two larger TTL values, the load balancing error does
not increase with the TTL because the client requests arrive
too infrequently to be affected by DNS caching. 

\textbf{Hidden clients:}
We also investigate the effect of the number of 
clients per LDNS. We keep the number of clients constant
across all LDNS servers. In order to keep the aggregate 
load fixed, we adjust the number of active LDNS servers.
We also keep the client idle period at 14s (mean)
so that in the case of a single client per LDNS, the
client request rate is higher than the advertised TTL
(15s).

\begin{figure}
\begin{center}
\includegraphics[width=0.45\textwidth]{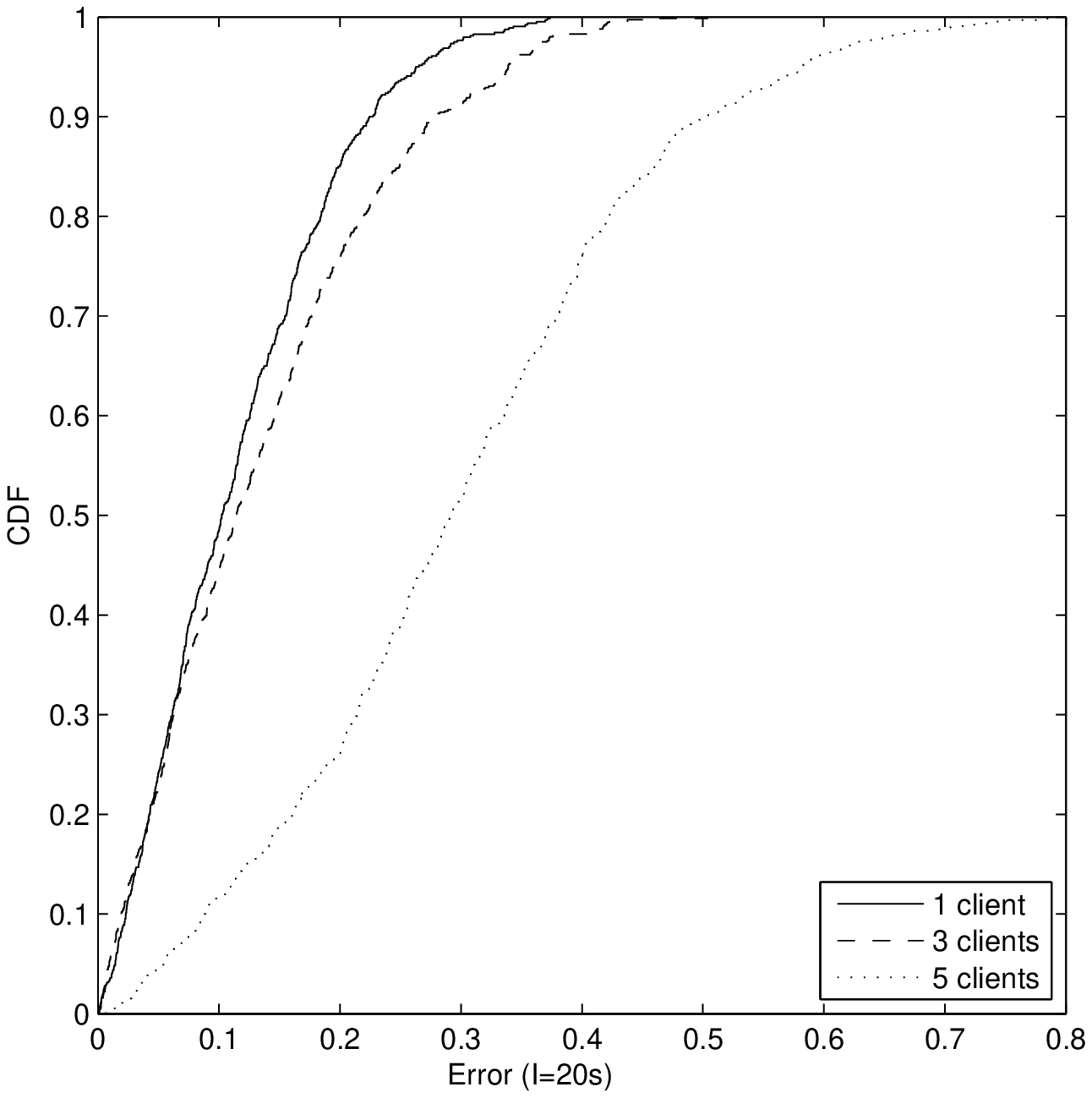}
\end{center}
\caption{Effect of hidden clients (RR).\label{fig:hideclients}}
\end{figure}

Figure~\ref{fig:hideclients} shows the load balancing
error when we increase the number of clients per LDNS 
from one to five. We see that the case for five clients
per LDNS shows a clear increase in the load balancing error 
compared to the case of one and three hidden clients.
Hidden clients increase load balancing errors
because the amount of traffic that corresponds to
each LDNS request grows with their count. 

%\begin{figure*}[t]
%\begin{minipage}[b]{0.48\linewidth}
%\begin{center}
%\includegraphics[width=1\textwidth]{figs/ttleffect.eps}
%\end{center}
%%\vspace{-20pt}
%\caption{Effect of advertised TTL (RR).\label{fig:ttl-effect}}
%\end{minipage}
%\begin{minipage}[b]{0.48\linewidth}
%\begin{center}
%\includegraphics[width=1\textwidth]{figs/hideclients.eps}
%\end{center}
%%\vspace{-20pt}
%\caption{Effect of hidden clients (RR).\label{fig:hideclients}}
%\end{minipage}
%%\vspace{-10pt}
%\end{figure*}

%\textbf{Summary:}
%Let us summarize the key observations from this section,
%in which we evaluated a round-robin DNS-ITE algorithm.
%The key parameters that affect the load balancing error
%are the aggregate load, the transfer size and its distribution,
%the number of hidden clients behind each LDNS server, 
%and, under certain conditions, the advertised/honored TTL value. 
%A content provider cannot control all these parameters.
%It can advertise a lower DNS TTL value, but that may be of limited
%value when the remote LDNS servers do not honor those
%advertisements.
%The aggregate load and the number of clients per LDNS server
%are not controlled by the content provider,
%while the transfer size distribution is also hard to control
%in practice.
%In the following section, we will examine whether it is 
%possible to further decrease the load balancing errors
%using a measurement-based algorithm.

%% file: mbscheme.tex
\section{Measurement-based DNS-ITE}
\label{sec:mbscheme}

In this section, we evaluate the performance of
a measurement-based load balancing scheme,
and examine the impact of the measurement history on
its accuracy.
Intuitively, the worst-case scenario for a round-robin
scheme is when requests for file sizes arrive in \{large, small\}
pairs.
%, and we end up sending all large transfers to one
%interface and all small transfers to the other. 
A measurement-based
scheme can alleviate such problems by routing clients 
based on the current link loads. 

In the measurement-based (MB) scheme that we consider in this
paper, we measure the incoming and outgoing load 
on each link of our content server 
using a sliding-window of duration $W$. On a new LDNS
request, we compare the latest utilization measurements of
the two links and advertise the interface with the minimum load.

\begin{figure}[t]
\begin{center}
\includegraphics[width=0.45\textwidth]{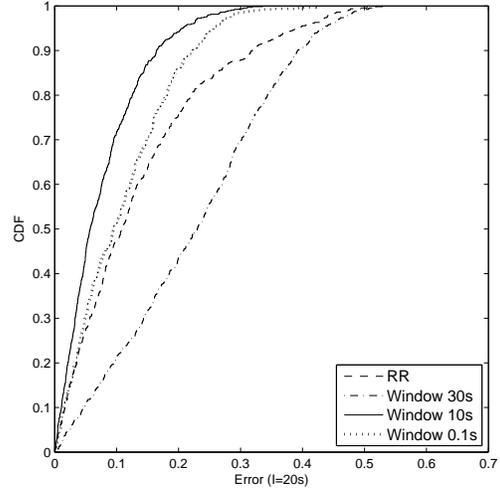}
\end{center}
\caption{RR versus MB load balancing.\label{fig:wndsize}}
\end{figure}

\textbf{Comparison between RR and MB:}
We compare MB with RR load balancing
using the same workload model we used
in the previous section. 
Figure~\ref{fig:wndsize} shows the distribution of
load balancing errors in the $I=20$s timescale with 
different window sizes, and with the RR scheme. 
We see that a large window of $W=30$s
performs worse than the RR scheme, while a small window
of $W=100$ms performs marginally better than RR. A
window size of $W=10$s is significantly better than RR.
It is clear that the parameter $W$ has a significant
impact on the accuracy of MB, and so we need to further understand why.

An important issue in any load balancing
scheme that relies on historical data 
is whether those past measurements are {\em stale},
meaning that they no longer reflect the current loads
\cite{mitzenmacher2000uoi}.
Intuitively, a larger value of $W$ is more prone
to such errors than a shorter $W$.
However, the magnitude of $W$ also controls
the {\em variance} of measurements.
A shorter $W$ introduces more noisy measurements,
making it harder to accurately estimate the load on each link.
An appropriate value of $W$ needs to consider 
carefully the staleness-vs-variance trade-off based
on the dynamics and burstiness of the underlying traffic.
%These issues, relative to the selection of $W$,
%are further discussed next. 

\textbf{Staleness-vs-variance trade-off:}
To illustrate the effects of
measurement staleness and variance
on load balancing accuracy we start with an
experiment using two simple, synthetic traffic models.
In the first model (CBR), each flow has a constant size
and duration and the packets are transmitted periodically. 
Here, the only variability in the aggregate traffic
is due to flow start/finish events. 
In the second model, the traffic is generated from 
an aggregate of Pareto renewal processes (inter-packet gaps), 
causing significant variability in the packet inter-arrivals. 
There are no flow start/finish events, however, which 
means that there are no rapid changes in the arriving
traffic rate. 

\begin{figure}[!t]
\begin{center}
\includegraphics[width=0.45\textwidth]{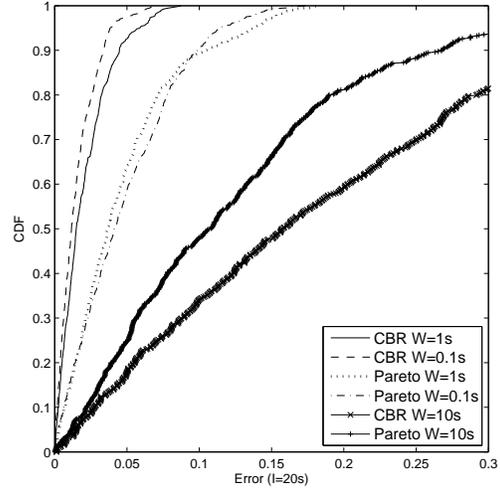}
\end{center}
\vspace{-0.2in}
\caption{Staleness-vs-variance trade-off.\label{fig:wnd-udp}}
\end{figure}

Figure~\ref{fig:wnd-udp} shows
the load balancing error distribution for the two models,
for different $W$s. For CBR, a smaller value of $W$ (0.1s)
performs best. The reason is that in CBR
the measurement variance is minimal (each flow consists of
periodic packet arrivals), and even a very short 
measurement window will suffice to estimate the load
of each link accurately. A larger $W$, say 10s,
is detrimental because it is subject to stale measurements 
(a different number of active flows than currently active).
With the Pareto model, on the other hand, we get lower
errors when $W$=1s compared to $W$=0.1s.
The reason is that this traffic is more bursty, and so
we need a longer time period in order to reliably 
know which link is more heavily loaded.

There is no ``optimal'' value of $W$ that is
independent of the statistical characteristics of the 
underlying traffic.
The general guideline that we can provide is that {\em $W$
should be as short as possible (to avoid the issue
of staleness) subject to the constraint that the  
measurement variance is sufficiently 
low to reliably show which link has the minimum load.} 
Note that the objective is not how to accurately estimate 
the load on each link. Instead, we are simply interested
in identifying the link with the minimum load.

%\begin{figure*}
%\begin{minipage}[b]{0.48\linewidth}
%\begin{center}
%\includegraphics[width=0.75\textwidth]{figs/wndsize-udp.eps}
%\end{center}
%%vspace{-20pt}
%\caption{Staleness-vs-variance trade-off.\label{fig:wnd-udp}}
%\end{minipage}
%\begin{minipage}[b]{0.48\linewidth}
%\begin{center}
%\includegraphics[width=0.75\textwidth]{figs/wndsize.eps}
%\end{center}
%%vspace{-20pt}
%\caption{RR versus MB load balancing.\label{fig:wndsize}}
%\end{minipage}
%%vspace{-15pt}
%\end{figure*}

\begin{figure}[t]
\begin{center}
\includegraphics[width=0.45\textwidth]{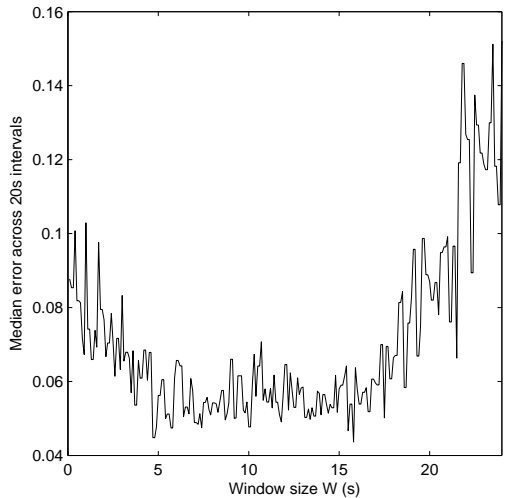}
\end{center}
\vspace{-0.2in}
\caption{Effect of $W$ on MB accuracy.\label{fig:wndsize-sims}}
\end{figure}

What is an appropriate value of $W$ for the TCP-based
workload that we used in our experiments?
To answer this question,
we simulated the MB load balancer on
packet traces from the $W=10$s experiment.
Specifically, we partitioned the experiment duration into 100ms intervals, 
identifying the load from different LDNS servers in each interval.
We then used a (hypothetical) window size $W$ to make MB load balancing
decisions for any new DNS requests in that interval,
and measured the resulting load balancing errors at $I=20$s. 
Figure~\ref{fig:wndsize-sims} shows the median error observed 
with $W$. We see that 
a measurement window $W$ in $[5,15]$s gives the lowest errors. 
With smaller window sizes we see the negative effects of variance, 
while with larger windows we observe the negative effects of staleness.
It is interesting that there is a wide range of $W$ in which the
load balancing accuracy is almost constant, implying that the 
selection of $W$ may not need to be fine-tuned in practice.

\textbf{Limitations of MB DNS-ITE:}
The MB scheme can improve load balancing accuracy
compared to RR, but it can still suffer from an 
intrinsic problem with DNS-based ITE:
there can be a significant delay between a DNS request
(and the associated load balancing decision) and the arrival of the actual
traffic that corresponds to that DNS request.
To illustrate this effect, consider the following simplified model
of DNS-ITE.

Suppose that we receive DNS requests at a constant rate $\lambda$. 
Traffic associated with each DNS request
can originate from multiple TCP connections and from multiple clients.
Further, there can be a significant delay between a 
load balancing decision %, upon the arrival of the DNS request,
and the arrival of the actual traffic load.
For a single connection, this delay includes the 
DNS response, the TCP connection establishment phase,
or the slow-start phase. % in which the connection grows its window.  
Let $\delta$ be the delay between a DNS request
and the time that the associated traffic arrives.
During that period, we can receive $\delta \lambda$ other
DNS requests, which will be routed without considering the load
that has been already committed (but not arrived) to each link. 
If $\delta > 1/\lambda$, i.e. if %the load initiation latency 
$\delta$ is significant compared to the inter-arrival time of
DNS requests, the MB scheme will not be able to correctly
amortize the load between the available links, at least in
short timescales. We view this as a fundamental problem
with the DNS-ITE method, which cannot be avoided given
that the delay $\delta$ is unknown and it varies across
LDNS servers and connection requests.

%% file: conclusion.tex
\section{Conclusion}

In this work, we looked at the problem of ingress
traffic load balancing using DNS-based techniques 
(ITE) in multihomed networks. We implemented
an ITE load balancer for a content server, and
designed a wide-area client testbed with realistic
workload characteristics.
% using our observations 
%of the popular GTLIB content distribution mirror.
Our contributions are two fold.

First, we showed that the 
accuracy of ITE can be impacted by factors
which include (1) aggregate client load, (2) DNS
 TTL policies in client networks, (3) 
hidden clients, and (4) heavy-tailed content 
sizes. We found that large aggregate load (1) can 
improve accuracy, while TTL violations (2), 3, 
and 4 can degrade performance. These observations
can be used to design a content distribution service
which aids load balancing.

Second, we showed that measurement-based (MB) schemes
improve performance over a round-robin scheme
when the length of measurement history is short
enough. We evaluated the impact of high 
variance and staleness in measurement history.
We finally looked at limitations of MB schemes due
to inherent nature of the ITE problem.